\input{epsf}
\documentstyle[prd,aps,twocolumn]{revtex}

\catcode`\@=11

\def\maketitle2{\par 
\begingroup
\let\cite\@bylinecite
\def\thefootnote{\fnsymbol{footnote}}%
\twocolumn[\@maketitle2\vskip2pc]%
\thispagestyle{plain}\@thanks
\endgroup
\def\thefootnote{\arabic{footnote}}%
\setcounter{footnote}{0}%
\let\maketitle2\relax \let\@maketitle2\relax
\let\@thanks\relax \let\@authoraddress\relax \let\@title\relax
\let\@date\relax \let\thanks\relax \let\@abstract\relax 
\let\@pacs\relax}

\def\abstract#1{\gdef\@abstract{{\par 
\bgroup
\ifdim\prevdepth=-1000pt \prevdepth0pt\fi
\hsize\columnwidth
\dimen0=-\prevdepth \advance\dimen0 by17.5pt \nointerlineskip
\small\vrule width 0pt height\dimen0 \relax}{~~}#1\egroup}}

\def\pacs#1{\gdef\@pacs{{\par 
\bgroup
\hsize\columnwidth \parindent0pt
\ifdim\prevdepth=-1000pt \prevdepth0pt\fi
\dimen0=-\prevdepth \advance\dimen0 by20pt\nointerlineskip
\egroup} PACS numbers:~#1}}

\def\@maketitle2{
\@preprint
\@title
\ifdim\prevdepth=-1000pt \prevdepth0pt\fi
\@authoraddress
\@date
\begin{list}{}{\leftmargin=0.10753\textwidth \rightmargin=\leftmargin
\itemsep=1pc\partopsep=-1pc}
\item\@abstract
\item\@pacs
\end{list}
}

\catcode`\@=12

\begin{document}
\draft
\preprint{LA-UR-98-2348}
\def \G {\Gamma}
\def\half{{1\over 2}}
\def \e {\eta }
\def \n {\nu}
\def \m {\mu}
\newcommand{\sq}{\lower.25ex\hbox{\large$\Box$}}

\title{Does Inflationary Particle Production suggest $\Omega_m < 1 $~?}
\author {Varun Sahni$^{1,*}$ and Salman Habib$^{2,\dagger}$}
\address {$^1$Inter-University Centre for Astronomy \& Astrophysics,
Post Bag 4, Pune 411007, India} 
\address{$^2$T-8, Theoretical Division, MS B285, Los Alamos National
Laboratory, Los Alamos, New Mexico 87545} 

\date{\today}

\abstract
{We study a class of FRW spacetimes with a nonminimally coupled light
massive scalar field. Values of the coupling parameter $\xi<0$ enhance
long range power in the vacuum expectation value of the
energy-momentum tensor $\langle T_{\mu\nu}\rangle$ and fundamentally
alter the nature of inflationary particle production: the energy
density of created particles behaves like an effective cosmological
constant, leading generically to $\Omega_m < 1$ in clustered matter
and providing a possible resolution of the ``$\Omega$ problem'' for
low density cosmological models.}

\pacs{04.62.+v, 98.80.Cq}

\maketitle2
\narrowtext

In this Letter we discuss particle creation of light nonminimally
coupled scalar fields due to the changing geometry of a spacetime
which underwent an early inflationary phase. Nonminimally coupled
fields have arisen in diverse cosmological contexts, {\em
e.g.}, density perturbations from strongly coupled scalars
\cite{denpert}, the possibility that a very light scalar particle such
as the axion may couple nonminimally to gravity \cite{tw88} and
production of primordial magnetic fields from nonminimal
electromagnetism \cite{magprl}. Ultra-light scalars (with $m
\sim H$) have been discussed in the context of pseudo-Nambu-Goldstone
bosons \cite{frieman95}.

As shown below, for negative values of the coupling the dominant
contribution to both $\langle \Phi^2\rangle$ and $\langle
T_{\mu\nu}\rangle$ comes from modes having wavelengths larger than the
Hubble radius which makes both quantities formally infrared (IR)
divergent. IR divergences cannot be renormalized away, instead a
physically motivated IR cutoff has to be invoked \cite{fp77}. The
IR-regularized $\langle T_{\mu\nu}\rangle$ describing particle
creation is finite, and behaves like an effective cosmological
constant as the Universe expands, consequently the energy density of
created particles can dominate the matter density leading to $\Omega_m
< 1$ in a flat Universe.

We consider a spatially flat FRW model with expansion factor either de
Sitter-like $a \propto \hbox{e}^{Ht}$, or power law $a \propto t^p$. 
Massive free scalar fields satisfy the wave equation
\begin{equation}
\lbrack\sq + \xi R + m^2\rbrack\Phi = 0
\label{eq:a2} 
\end{equation}
where $R$ is the Ricci scalar and $\xi$ parametrizes the coupling to
gravity, $\xi = 0,~1/6$ corresponding to minimal and conformal
coupling respectively. In a spatially flat FRW Universe the field
variables separate and $\Phi_k = (2\pi)^{-3/2}\phi_k (\eta) \,
e^{-i{\bf k}\cdot{\bf x}}$. The comoving wavenumber $k = 2\pi
a/\lambda$ where $\lambda$ is the physical wavelength of scalar field
quanta. Defining the conformal field $\chi_k=a\phi_k$ and using $R =
6{\ddot a}/a^3$ (differentiation being with respect to the conformal
time $\eta = \int dt/a$), Eq. (\ref{eq:a2}) leads to,
\begin{equation}
\ddot \chi_k + [k^2 + m^2a^2 - (1 - 6\xi){\ddot a}/a]\chi_k = 0~.
\label{eq:a4}
\end{equation}
For de Sitter space (\ref{eq:a4}) has the exact solution
\begin{equation}
\chi_k(\eta)=c_1 \sqrt{\eta}H_{\mu}^{(2)}(k\eta) + c_2
\sqrt{\eta}H_{\mu}^{(1)}(k\eta)~,
\label{eq:a5}
\end{equation}
where $\mu^2 = 9/4 -12\xi -m^2/H^2$, $c_1=\sqrt{\pi}/2,~c_2=0$ gives
the state associated with the Bunch-Davies vacuum \cite{bd78}.

For ultra-light fields $m/H \sim 1$ today, corresponds to $m/H \sim
10^{-60}$ during inflation, hence the field modes can be treated as
being effectively massless. Setting $m = 0$ in (\ref{eq:a4}) and
specializing to a power law expansion
\begin{equation}
a = (t/t_0)^p \equiv\left({\eta\over \eta_0}\right)^{(1-2 \nu)/2}~,
\end{equation}
where $2\nu = (1-3p)/(1-p)$, (the inflationary range $p > 1$
corresponds to $\nu \ge 3/2$ and for $a\propto\hbox{e}^{Ht}$, $\nu =
3/2$), we find exact solutions of (\ref{eq:a4}) having the form
(\ref{eq:a5}) with $\mu^2-1/4=(\nu^2-1/4)(1-6\xi)$. For $\xi=1/6$,
$\mu = 1/2$ while for $\xi=0$, $\mu = \nu$. The choices
$c_1=\sqrt{\pi}/2,~c_2=0$ correspond to the adiabatic vacuum state.

To study quantum fluctuations we define the operator
\begin{equation}
\Phi(x) = \int d^3k\left[a_k\Phi_k({\bf x},\eta)
+ a_k^{\dagger} \Phi_k^*({\bf x},\eta)\right]
\label{eq:a6}
\end{equation}
where $a_k,~a_k^{\dagger}$ are annihilation and creation operators
$[a_k,a_{k'}^{\dagger}] = \delta_{kk'}$, defining the vacuum state
$a_k|0\rangle = 0$ $\forall k$. The two-point function
\begin{equation}
\langle \Phi(x)\Phi(x')\rangle_{vac}={1\over (2\pi)^3}\int d^3k
\hbox{e}^{i{\bf k}\cdot({\bf x}-{\bf x}')}\phi_k(\eta)\phi_k^*(\eta')
\label{twopt}
\end{equation}
is IR divergent over a certain range of $\mu$ values. To see this,
substitute Eq. (\ref{eq:a5}) in Eq. (\ref{twopt}), using
$\phi_k=\chi_k/a$ and the small-argument limit of the Hankel functions
\begin{equation}
H_\mu ^{(2,1)}(k\eta){\mathrel{\mathop{\simeq}_{k\eta < 1} } }
{({k\eta/2})^\mu \over \Gamma
(1+\mu)} \pm {i\over \pi} \Gamma (\mu) \left({k \eta \over
2}\right)^{-\mu}~.
\label{eq:a8}  
\end{equation}
The integral controlling the presence of IR divergences is  
\begin{equation}
\int dk k^2 k^{-2\mu}|c_1 - c_2|^2~.
\label{irint}
\end{equation}
For the adiabatic vacuum state (\ref{eq:a5}), IR divergences arise
when $\mu^2 \ge 9/4$. For $\xi=0$, IR divergences are encountered for
$p\ge 2/3~(p \ne 1)$ \cite{fp77}. The situation is substantially
different for $\xi\neq 0$ (Fig. 1), IR divergences being present over
a wide range of expansion rates: $p<0$; $p>1/2~(p\neq 1)$;
$0<p<1/2$. Since there is no particle production when
$p=0,~1/2,~\xi=1/6$, these special cases are free of IR divergences.
 
\vspace{.4cm}
\epsfxsize=6cm
\epsfysize=4.5cm
\centerline{\epsfbox{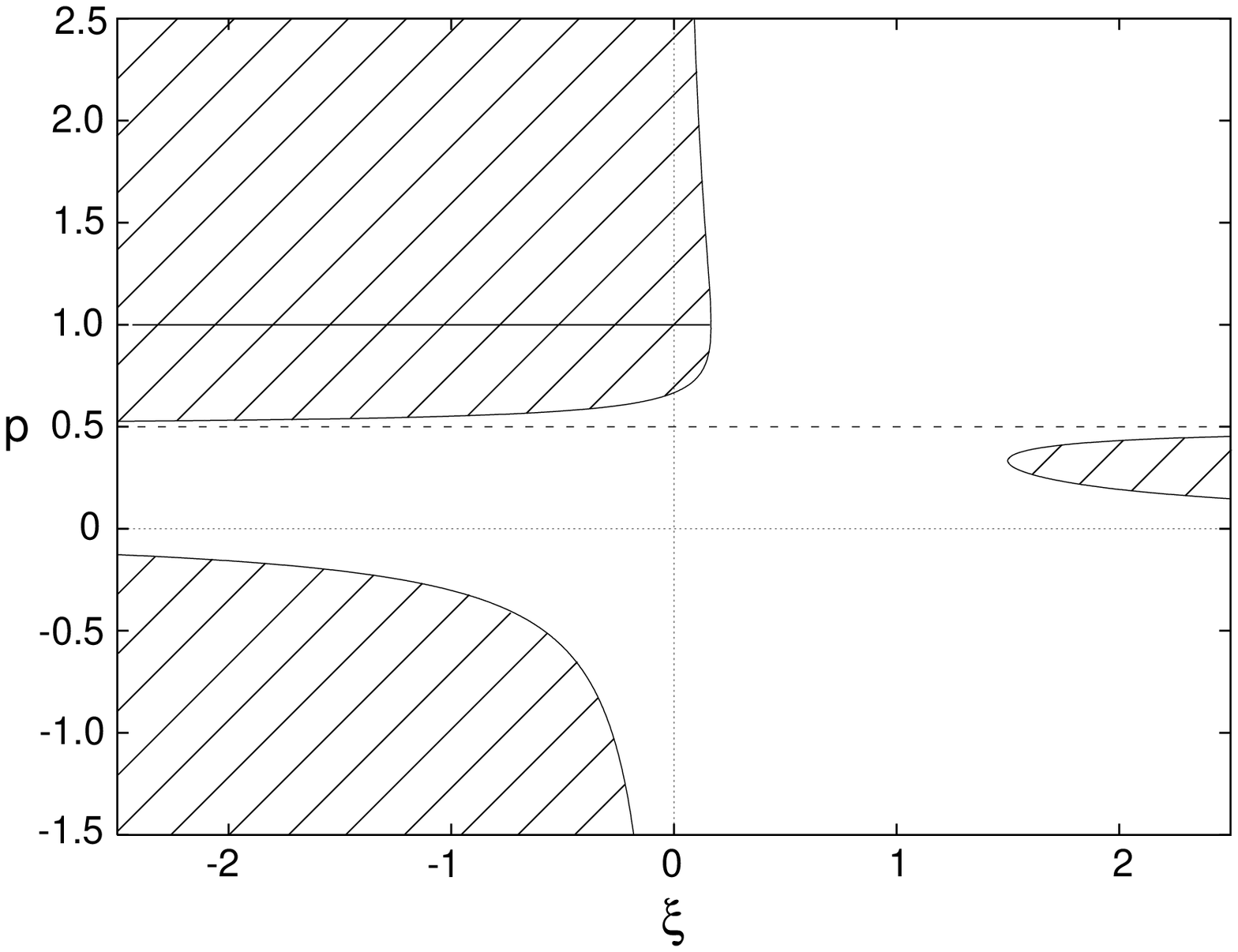}}
\vspace{0cm}
{FIG. 1. {\small{IR divergent regions (shaded) of
$\langle\Phi^2\rangle$: The special cases $p = 0,~1/2,~1$ and
$1/6\leq\xi\leq 3/2$ have no IR divergences.}}} \\  

Significantly, $\langle T_{\mu\nu}\rangle$ can also be IR divergent:
The (bare) energy density in a spatially flat Universe is
\begin{eqnarray}
& \rho  & = \langle T_{00}\rangle = {1\over 4\pi^2a^2}\int_0^\infty dk
k^2 \left(|\dot\phi_k|^2 + k^2|\phi_k|^2\right) 
\nonumber\\ 
& &+ {3\over 2\pi^2}\xi H\int_0^\infty dk 
k^2\left[H~|\phi_k^2|+{\phi_k \dot
\phi_k^*+\dot\phi_k\phi_k^*\over a}\right]  \nonumber\\
& & + \frac{m^2}{4\pi^2}\int_0^\infty dk k^2 |\phi_k|^2~.
\label{eq:a11} 
\end{eqnarray}
Restricting attention momentarily to $\xi = 0$, $m = 0$, we conclude
that $\langle T_{00}\rangle$ is IR divergent for $|\nu| \ge 5/2$, {\em
i.e.}, $3/4 \le p \le 2 ~~(p \ne 1)$. [We correct a minor error in
Ref. \cite{fp77} who quote $2/3 \le p \le 2 ~(p \ne 1)$.] This is a
much smaller range than for the two-point function which is divergent
for $p \ge 2/3 ~(p \ne 1)$. In contrast, the key aspect of nonminimal
coupling is that $\langle T_{00}\rangle$ contains terms proportional
to $\xi\langle\Phi^2\rangle$, and consequently (for ultra-light
fields) $\langle T_{00}\rangle$ is IR divergent {\em over the same
range of parameters} as $\langle\Phi^2\rangle$.

The curing of IR divergences requires that mode functions be modified
in the IR limit. The only freedom to accomplish this in
Eq. (\ref{irint}) is to change the behavior of $|c_1-c_2|$ as
$k\rightarrow 0$: There are no IR divergences if $\lim_{k \rightarrow
0}|c_1 - c_2|^2 \propto O(k^{2|\mu|-2})\rightarrow 0$ (maintaining
$c_2 = 0$, $c_1 = 1$ at large $k$). One way to determine $c_1$ and
$c_2$ is to assume the existence of a pre-inflationary
radiation-dominated phase \cite{fp77,vf82}, as a result
\cite{sahni88}, $|c_1 - c_2|^2 \simeq [1 + 4\pi^2 C
(k\tilde{\eta}_0)^{1-2|\mu|}]^{-1}$, where $\tilde{\eta}_0$ marks the
onset of inflation. Finiteness in the ultraviolet is achieved by
imposing a cutoff at the horizon scale. It then follows that
\begin{eqnarray}
\langle\Phi^2\rangle = \bar{C}\eta^{2(\nu -
|\mu|)}\int_{\tilde{\eta}_0^{-1}}^{\eta^{-1}} dk k^2 k^{-2|\mu|}
\label{eq:d5}
\end{eqnarray}
where $\bar{C} = C\tilde{\eta}_0^{1-2\nu}$. The behavior of
$\langle\Phi^2\rangle$ depends crucially upon the value of $\nu -
|\mu|$. For $\xi=0$, $\langle\Phi^2\rangle ~\simeq
\bar{C}\int_{1/\tilde{\eta}_0}^{1/\eta} dk~k^{2(1-\nu)}$, which gives
for exponential inflation ($\nu = 3/2$) the standard result
$\langle\Phi^2\rangle \simeq H^3~\Delta t/(4\pi^2)$ \cite{lin90}. In
the case of power law inflation $\nu > 3/2$, and
$\langle\Phi^2\rangle$ freezes to a large value at late times
\cite{sahni88,pf88}. With negative values of $\xi$, $\mu > \nu$,
$\langle\Phi^2\rangle$ grows with time approaching the asymptotic
form:
\begin{equation}
\langle\Phi^2\rangle \, { \mathrel{\mathop{=}_{\eta \rightarrow 0}}}
\,~ {C\over 2|\mu| -
3}~\left({\tilde{\eta}_0\over\eta}\right)^{2(|\mu| -
\nu)}\tilde{\eta}_0^{-2}  
\label{eq:d5a}
\end{equation}
at late times. 
For $\xi \ll -1/6$ and $\nu^2 \gg 1/4$  we have $|\mu|
\approx \nu\sqrt{6|\xi|}$, which substituted in (\ref{eq:d5a}) gives
\begin{equation}
\langle\Phi^2\rangle ~\propto a^c, ~~ c =
\frac{4\nu}{2\nu-1}(\sqrt{6|\xi|} - 1) > 1 
\end{equation}
Thus $\xi <0$ can greatly accelerate the growth of fluctuations in
inflationary models. Even for minimal coupling, $\xi = 0$, IR finite
physical quantities, {\em e.g.}, $\langle\Phi(x)\Phi(x')\rangle$ and
$\langle T_{00}\rangle$, remain sensitive to the presence of long
range power in the field modes. This is reflected in the growth of
scalar fluctuations, generation of density fluctuations, and the
quantum creation of gravitational waves.

We now examine particle production in a Universe that inflates and
then transits to a matter dominated regime of expansion. To do this,
we return to Eq. (\ref{eq:a4}): this equation closely resembles the
one dimensional Schr\"{o}dinger equation, the role of the ``potential
barrier'' $V(x)$ being played by $V(\eta) = - m^2a^2 + (1 -
6\xi){\ddot a}/a$ \cite{ss92}. The form of the barrier is shown in
Fig. 2 assuming $\xi < 1/6$, $m \simeq 0$.  The process of
super-adiabatic amplification of zero-point fluctuations (particle
production) can be qualitatively described as follows: the amplitude
of modes having wavelengths smaller than the Hubble radius decreases
conformally with the expansion of the Universe, whereas that of
larger-than Hubble radius modes freezes (if $\xi = 0$) or grows with
time ($\xi < 0$). Consequently, modes with $\xi \le 0$ have their
amplitude super-adiabatically amplified on re-entering the Hubble
radius after inflation (Fig. 2).

The non-vacuum state of the scalar field at late times ($\eta >
|\eta_0|$) is described by a linear superposition of positive and
negative frequency solutions
\begin{equation}
\phi_{out}(k,\eta) = \alpha\phi_k^+ + \beta\phi_k^-,
\label{eq:ab2}
\end{equation}
where $\phi_k^{+,-} = (\sqrt {\pi
\eta_0}/2)(\eta/\eta_0)^{\bar{\nu}}H_{|\bar{\mu}|}^{(2,1)}
(k\eta)$. [$a \propto t^p, ~p < 1$; $2\bar\nu = (1-3p)/(1-p)$, $-3/2
\le \bar{\nu} \le -1/2 \Longleftrightarrow$ $2/3 \ge p \ge 1/2$
corresponding to matter with an equation of state $0 \le w \le 1/3$,
$\bar{\mu}^2-1/4=(\bar{\nu}^2-1/4)(1-6\xi)$.] The transition from
inflation to a radiation/matter dominated epoch is marked by $\eta_0$,
$H_{rh}^2 \equiv 1/\eta_0^2$ being the Hubble parameter at
reheating. [We assume $m^2 < |\xi| R$ or $m^2/H^2 < 6|\xi|({\bar\n}^2
- 1/4)/({\bar\n} - 1/2)^2$ allowing us to treat field modes as being
effectively massless.]

The Bogoliubov coefficients $\alpha$ and $\beta$ are determined by
matching $\phi_{in}^+(k,\eta),~\dot\phi_{in}^+(k,\eta)$ given in 
(\ref{eq:a5}) and $\phi_{out}(k,\eta),~\dot\phi_{out}(k,\eta)$ at
$\eta=\eta_0$. For modes with $k\eta_0 < 1$ we obtain 
\begin{eqnarray}
\alpha + \beta & = &
A\frac{i}{\pi}\Gamma(\m)\Gamma(1+\bar\m)
\left(\frac{k\e_0}{2}\right)^{-(\m+\bar\m)}\nonumber\\
&& + B\frac{\G(1+\bar\m)}{\G(1+\m)}
\left(\frac{k\e_0}{2}\right)^{\m-\bar\m}\nonumber\\ 
\alpha - \beta & = & C {i\pi\over \G(\bar\m)\G(1 +
\m)}\left(\frac{k\e_0}{2}\right)^{\m+\bar\m}\nonumber\\
&& + D \frac{\G(\m)}{\G(\bar\m)}
\left(\frac{k\e_0}{2}\right)^{\bar\m-\m}\nonumber\\ 
|\alpha|^2 - |\beta|^2 & = & 1
\label{eq:ab4}
\end{eqnarray}
where, $A= (\bar\m + |\bar\n| +\n - \m)/2\bar\m$,
$B = (\bar\m + |\bar\n| + \n + \m)/ 2\bar\m$,
$C = (\n + \m + |\bar\n| - \bar\m)/2\bar\m$,
$D = (\bar\m - |\bar\n| + \m - \n)/2\bar\m$.
(Indices $\nu, \mu$ refer to the inflationary epoch; 
$\bar\nu, \bar\mu$ to the matter/radiation dominated epoch.)

\vspace{.4cm}
\epsfxsize=7.5cm
\epsfysize=6.5cm
\centerline{\epsfbox{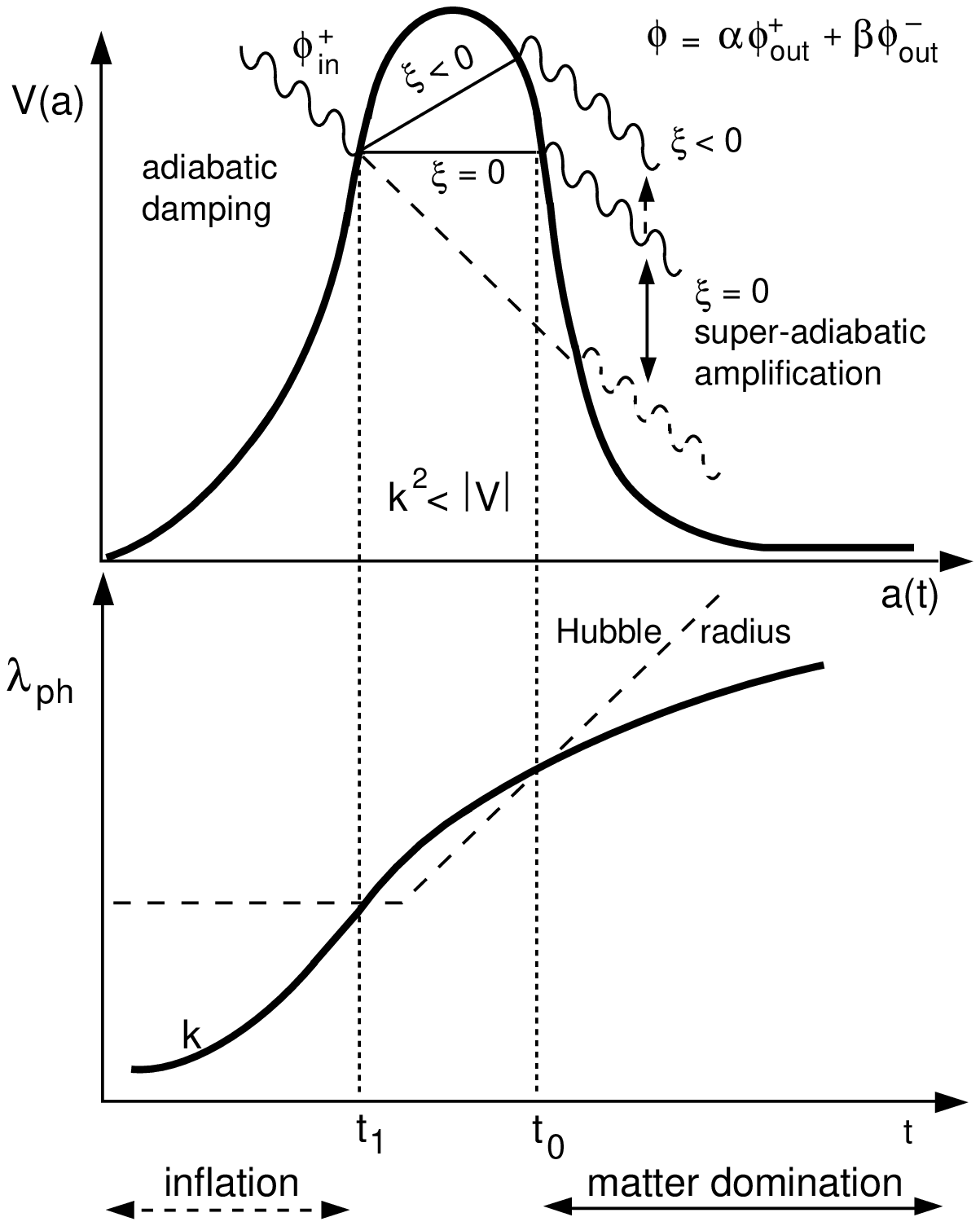}}
\vspace{.35cm}

{FIG. 2. {\small{The time-like ``potential barrier'' $V(\eta)$ for
inflation followed by a matter dominated epoch. 
}}}\\

Since $|\beta|^2 \propto (k\eta_0)^{-2(\m + \bar\m)}$ and
$\mu=\mu(\nu,\xi)$, the number density of created particles is
sensitive to: 1) the inflationary expansion rate parametrized by
$\nu$, 2) the equation of state after inflation, parametrized by
$\bar\n$, 3) the coupling to gravity $\xi$. More particles are created
for $\xi < 0$ than for $\xi = 0$ (see Fig. 2 and Ref. \cite{tw88}).

From (\ref{eq:ab2}) and (\ref{eq:ab4}) we find that on larger than
horizon scales ($k\e < 1$), 
\begin{equation}
\phi_{out} \simeq i~A\sqrt{\e_0/4\pi}\G(\m)
(k\e_0/2)^{-\m}(\e/\e_0)^{\bar\m-|\bar\n|}  
\label{eq:ab6}
\end{equation}
from which follows the important observation that
$\langle\Phi^2\rangle_{out} = 1/2\pi^2\int dk|\phi_{out}|^2k^2$ has
{\em exactly the same} infrared properties as
$\langle\Phi^2\rangle_{in}$ and that on scales {\em larger} than the
horizon $\langle\Phi^2\rangle_{out}$ grows with time at a rate
determined by $\bar\m-|\bar\n| > 0$ ($|\bar\n| = 3/2$ in a matter
dominated Universe).  This result bears directly on the energy density
of created particles $\langle T_{00}\rangle_{out}$, which can be
determined from (\ref{eq:a11}) after substituting $\phi
\rightarrow \phi_{out}$.  Eq. (\ref{eq:a11}) informs us that the main
contribution to $\langle T_{00} \rangle_{out}$ comes from long
wavelength modes and, for $\xi < 0$, $\langle T_{00} \rangle_{out}$ is
IR divergent if the field is effectively massless. A radiation
dominated phase prior to inflation leads to an effective IR cutoff
$k_{min} \simeq \tilde{\e_0}^{-1}$ in $\langle T_{00} \rangle$, since
IR divergent states cannot arise from IR finite initial conditions
\cite{fulling}. In addition, a high frequency cutoff appears due to
suppression of particle creation at large $k$ \cite{bd82}. For
particles created during inflation, this cutoff is set by the Hubble
parameter at the end of reheating and the commencement of radiation
domination: $k_{max} \simeq H_{rh}\simeq \eta_0^{-1}$
\cite{bd82,bavs}.

The integration limits in (\ref{eq:a11}) are therefore
$\int_{k_{min}}^{k_{max}}$.  For $m/H \le 1$, $\langle T_{00} \rangle$
is dominated by modes larger than the Hubble radius, thus the
integration limits are effectively
$\int_{\tilde{\e_0}^{-1}}^{\e^{-1}}$ leading, for a matter dominated
Universe, to
\begin{equation}
\langle\Phi^2\rangle \simeq \frac{N}{\e_0^2}
\left(\frac{\e}{\e_{MD}}\right)^{2\bar\m - 3} 
\left(\frac{\tilde\e_0}{\e_0}\right)^{2\m-3}\left[1 -
\left(\frac{\e}{\tilde\e_0}\right)^{2\m-3}\right]
\label{eq:ab7}
\end{equation}
where $N = (A^2/8\pi^3)2^{2\m}\G^2(\m)/(2\m-3)$.

Although a full treatment of the semiclassical Einstein equations
$G_{\mu\nu} = - 8\pi G(T_{\mu\nu} + \langle T_{\mu\nu}\rangle)$ lies
beyond the scope of this work, it is easy to perform a qualitative
analysis.  For small values $|\xi| < 1$, the term $1/(4\pi a^2)\int dk
k^4 |\phi_k|^2$ in Eq. (\ref{eq:a11}) is dominated by horizon-size
modes, and is small compared to the remaining terms in $\langle
T_{00}\rangle$, which are dominated by the larger infrared cutoff
scale $\tilde{\e_0}$ ($\tilde{\e_0}\gg\eta$). These terms, excluding
$m^2\langle\Phi^2\rangle$, are of the form $H^2\langle\Phi^2\rangle$
and can be absorbed into the left hand side of the ($00$) Einstein
equation leading to:
\begin{equation}
3H^2 \simeq 8\pi \bar G\big\lbrack \rho_m + \frac{1}{2}m^2
\langle\Phi^2\rangle\big\rbrack
\label{eq:ab8} 
\end{equation}
where $\bar G \simeq G/(1 + 8\pi G|\xi|\langle\Phi^2\rangle)$.  For
ultra-light fields $|\xi| < 0.1$ ensures that $\langle\Phi^2\rangle$
grows at a slow rate, satisfying constraints on the time variation of
$\bar G$ \cite{will}. Consequently,
\begin{equation}
8\pi \bar G\langle\Phi^2\rangle \simeq N
\left(\frac{H_{rh}}{m_{pl}}\right)^2(1 + z_{MD})^{\m + \bar\m -
3}\left(\zeta\frac{T_{rh}}{T_{MD}}\right)^{2\m - 3} 
\end{equation}
where $\zeta = \tilde\e_0/\e$.  (Note that $\zeta \equiv
k_*/k_{min} = \exp{\int_{t_{in}}^{t_*} H dt}$, where $t_{in}$ is the
beginning of inflation, and $t_*$ the time when a mode entering the
horizon today, left the Hubble radius during inflation.  In general
$\zeta$ can be very large, $\log{\zeta} \gg 1$.)  Substituting
$H_{rh}/m_{pl} \simeq 10^{-5}$, $T_{rh} \simeq 10^{14}~GeV$, $T_{MD}
\simeq (1 + z_{MD})\times 2.3\times 10^{-13}~GeV$, $1 + z_{MD} \simeq
23,219~\Omega_m h^2$, we find that $8\pi \bar{G} \langle\Phi^2\rangle$
can also be very large (since $\bar{G} < G$ this holds for $8\pi G
\langle \Phi^2\rangle$ as well). For large $8\pi
G|\xi|\langle\Phi^2\rangle \gg 1$, $\bar G \simeq
1/(8\pi|\xi|\langle\Phi^2\rangle$) and
\begin{eqnarray}
\Lambda_{eff} \equiv & 8\pi\bar
G\langle T_{00}\rangle \simeq m^2/2|\xi|\nonumber\\
\Omega_{\phi}
\equiv & \Lambda_{eff}/3H^2 \simeq \frac{1}{6|\xi|}(m/H)^2
\label{omega}
\end{eqnarray}
We thus find that the energy density of created particles behaves like
an {\em effective cosmological constant}. Furthermore, for ultralight
fields $m/H \sim 1$, $\Omega_{\phi}$ can be a significant fraction of
the total density of the Universe at late times. (Recent observations
of high redshift supernovae suggest $\Omega_{\Lambda} > 0$, which
could be quite large if the Universe is flat \cite{perl98}).

At first glance it may appear strange that the equation of state of
created nonminimal scalars is $p \simeq-\rho$ and not the familiar $p
= \rho/3$ expected for relativistic particles. This is because the
dominant contribution to $\langle T_{00}\rangle$ comes from
wavelengths larger than the horizon size, as a result $\langle
T_{\m\n}\rangle\simeq\frac{1}{2} g_{\m\n}m^2\langle\Phi^2\rangle$ at
late times.  In a similar context it is well known that $\langle
T_{00}\rangle$ for gravitational waves (and minimally coupled massless
scalars) created during inflation is dominated by wavelengths of order
the horizon size, because of which the equation of state of relic
gravitational waves is identical to that of classical matter driving
the expansion, and in a matter-dominated Universe, is $p=0$
\cite{bavs}. (For wavelengths larger than the Hubble radius the
distinction between ``real'' particles and vacuum polarization becomes
ambiguous.)

Our analysis indicates that, at a local level, the Universe may be
filled with a quasiclassical, stochastic, nearly homogeneous
nonminimal scalar field having a Gaussian probability
distribution with dispersion $\langle \Phi^2\rangle$. (This
description agrees with the general stochastic approach to
inflationary cosmology \cite{lin90,star86}.)  Due to cosmic variance,
the local value of $\Phi^2$ will generally not equal $\langle
\Phi^2\rangle$, remarkably this does not affect the value of
$\Omega_\phi$ since (\ref{omega}) does not depend upon $\Phi$
(provided $\Phi$ is large).

To summarize, the inflationary production of light nonminimal
particles ($\xi < 0$) leads to an energy density which mimics an
effective cosmological constant implying $\Omega_m < 1$ in clustered
matter in a critical density Universe. Hence, a low density Universe 
can exist without an ``$\Omega$ problem'' making inflation
compatible with observations indicating a low value for $\Omega_m$
\cite{coles94,ostr95}. Since nonminimal scalars naturally arise in 
the low energy limit of string theory \cite{dv96}, our results may 
apply to a wider class of models than the ones considered here.

It is a pleasure to thank Yuri Shtanov and Alexei Starobinsky for
stimulating discussions.

\end{document}